\documentclass[aps,prb,twocolumn,showpacs,superscriptaddress,groupedaddress,amsmath,amssymb]{revtex4}  

\usepackage{graphicx}
\usepackage{dcolumn}
\usepackage{bm}
\usepackage{ulem}
\usepackage{color}

\begin{document}


\title{Flux qubit noise spectroscopy using Rabi oscillations under strong driving conditions}

\author{Fumiki Yoshihara}
\email{fumiki.yoshihara@riken.jp}
\affiliation{Center for Emergent Matter Science (CEMS), RIKEN, Wako, Saitama 351-0198, Japan}

\author{Yasunobu Nakamura}
\affiliation{Center for Emergent Matter Science (CEMS), RIKEN, Wako, Saitama 351-0198, Japan}
\affiliation{Research Center for Advanced Science and Technology (RCAST), The University of Tokyo, Komaba, Meguro-ku, Tokyo 153-8904, Japan}

\author{Fei Yan}
\affiliation{Department of Nuclear Science and Engineering, Massachusetts Institute of Technology (MIT), Cambridge, Massachusetts 02139, USA}

\author{Simon Gustavsson}
\affiliation{Research Laboratory of Electronics, MIT, Cambridge, Massachusetts 02139, USA}

\author{Jonas Bylander}
\affiliation{Research Laboratory of Electronics, MIT, Cambridge, Massachusetts 02139, USA}
\affiliation{Dept. of Microtechnology and Nanoscience, Chalmers University of Technology, SE-412 96 Gothenburg, Sweden}

\author{William D. Oliver}
\affiliation{Research Laboratory of Electronics, MIT, Cambridge, Massachusetts 02139, USA}
\affiliation{MIT Lincoln Laboratory, 244 Wood Street, Lexington, Massachusetts 02420, USA}

\author{Jaw-Shen Tsai}
\affiliation{Center for Emergent Matter Science (CEMS), RIKEN, Wako, Saitama 351-0198, Japan}
\affiliation{NEC Smart Energy Research Laboratories, Tsukuba, Ibaraki 305-8501, Japan}

\date{\today}

\begin{abstract}
We infer the high-frequency flux noise spectrum in a superconducting flux qubit
by studying the decay of Rabi oscillations under strong driving conditions.
The large anharmonicity of the qubit and its strong inductive coupling to a microwave line enabled
high-amplitude driving without causing significant additional decoherence.
Rabi frequencies up to 1.7~GHz were achieved, approaching the qubit's level splitting of 4.8~GHz,
a regime where the rotating-wave approximation breaks down as a model for the driven dynamics.
The spectral density of flux noise observed in the wide frequency range decreases with increasing frequency
up to 300~MHz,
where the spectral density is not very far from the extrapolation of the $1/f$ spectrum obtained from
the free-induction-decay measurements.
We discuss a possible origin of the flux noise due to surface electron spins.

\end{abstract}


\maketitle
Flux noise has been investigated for decades to improve
stability and sensitivity in superconducting flux-based devices.
Its power spectral density (PSD) has been studied
in superconducting quantum interference devices (SQUIDs)~\cite{Koch83IEEE, Wellstood87APL}
and in various types of superconducting qubits,
such as charge,~\cite{Ithier05PRB}
flux,~\cite{Fum06PRL, Kakuyanagi07PRL, Lanting09PRB, Harris08PRL, Harris10PRB, Bylander11NatPhys, YanFei2012PRB}
and phase qubits.~\cite{Claudon06PRB, Bialczak07PRL, Bennett09, Sank12PRL}
The spectra typically follow 1/$f$ frequency dependence with a spectral density of
1--10~$\mu \Phi_0/\sqrt{\rm Hz}$ at 1~Hz,
where $\Phi_0 = h/2e$ is the superconducting flux quantum. 
The accessible frequency range of the PSD was limited to approximately 10~MHz in spin-echo measurements~\cite{Fum06PRL, Kakuyanagi07PRL, Fum10PRB, Bylander11NatPhys}
and was extended to a few tens of megahertz using Carr--Purcell--Meiboom--Gill pulse sequences.~\cite{Bylander11NatPhys}
Recently, spin-locking measurements provided the PSD up to approximately 100~MHz,~\cite{YanFei2013NatCom}
and a study of qubit relaxation due to dressed dephasing in a driven resonator revealed the PSD
at approximately 1~GHz.~\cite{Slichter12PRL}
The spectrum in a higher-frequency range would give further information
for better understanding of the microscopic origin of the flux fluctuations.

Decay of Rabi oscillations has also been used as a tool to characterize the decoherence in superconducting qubits.
PSDs of fluctuating parameters, such as charge, flux, or coupling strength to an external two-level system,
at the Rabi frequency $\Omega_{\rm R}$ can be detected.~\cite{Martinis03PRB, Ithier05PRB, Bylander11NatPhys, Lisenfeld10PRB}
The Rabi frequency is proportional to the amplitude of the driving field
for weak to moderate driving at the qubit transition frequency,
and Rabi frequencies in the gigahertz range have been achieved under a strong driving field.~\cite{Yasu01PRL, Chiorescu04Nat, Saito06PRL}
However, the decay was not systematically studied because of the presence of extrinsic
decoherence mechanisms under the strong driving conditions.

To induce fast Rabi oscillations without significant extra decoherence,
we choose a flux qubit having strong inductive coupling to a microwave line and
large anharmonicity, $|(\omega_{12}-\omega_{01})/\omega_{01}|$,
to avoid unwanted excitations to the higher energy levels,
where $\omega_{ij}$ is the transition frequency between the $|i\rangle$ and $|j\rangle$ states.
We measured Rabi oscillations in a wide range of $\Omega_{\rm R}/2\pi$ from $2.7$~MHz to $1.7$~GHz,
and evaluated the PSD of flux fluctuations at each $\Omega_{\rm R}$.
The observed PSD decreases up to 300~MHz,
where the spectral density is approximately $10^{-20}\,(\Phi_0)^2\mathrm{rad}^{-1}\mathrm{s}$.
Above 300~MHz, the PSD scatters and slightly increases.
We discuss a possible origin of the flux fluctuations due to surface electron spins.

The Hamiltonian of a flux qubit with a flux drive and in the presence of fluctuations
can be written in the persistent current basis as
\begin{multline}\label{Hpc}
    \mathcal{H}_{\rm pc} = -\frac{\hbar}{2} [(\Delta\sigma_{x} + \varepsilon\sigma_{z})
    + \varepsilon_{\rm{mw}}\cos (\omega_{\rm mw} t) \,\sigma_z \\
    + \delta \Delta(t) \sigma_x
    + \delta \varepsilon(t) \sigma_z ],
\end{multline}
where $\sigma_{x}$ and $\sigma_{z}$ are Pauli matrices,
$\Delta$ is the tunnel splitting between two states with opposite persistent current direction
along the qubit loop $I_{\rm p}$,
and $\hbar\varepsilon = 2I_{\rm p}\Phi_0 n_{\phi}$ is the energy bias between the two states.
Here the flux bias through the loop $\Phi_{\rm ex}$ is normalized by $\Phi_0$ as
$n_{\phi}=\Phi_{\rm{ex}}/\Phi_0-0.5$.
The first term on the right-hand side of Eq.~(\ref{Hpc}) represents the flux qubit with a static flux bias.
The transition frequency can be written as $\omega_{01} = \sqrt{\Delta^2 + \varepsilon^2}$.
We find $\omega_{01}=\Delta$ and $\partial \omega_{01}/\partial n_{\phi}=0$ at $n_{\phi}=0$;
this is the optimal flux bias condition where dephasing due to fluctuations of $n_{\phi}$ is minimal.
The second term is an ac drive at frequency $\omega_{\rm mw}$ with the amplitude $\varepsilon_{\rm mw}$
to induce Rabi oscillations.
The third and fourth terms represent fluctuations of $\Delta$ and $\varepsilon$, respectively.
In the present sample, $\varepsilon$ is tunable via $n_{\phi}$
while $\Delta$ is fixed.

There exist a few dominant contributors to the decay of Rabi oscillations:
the quasistatic noise;
the noise at $\omega_{01}$, which causes the qubit energy relaxation;
and the noise at $\Omega_{\rm R}$.~\cite{Falci05PRL, Ithier05PRB, Bylander11NatPhys}
The resulting decay envelope $A_{\rm{env}}(t)$ is described as
\begin{eqnarray}\label{RabiDecay}
    A_{\rm{env}}(t) = A_{\rm{st}}(t) \exp (-\Gamma_{\rm R}^{\rm{exp}} t),
\end{eqnarray}
where $A_{\rm{st}}(t)$ is the contribution from the quasistatic noise, which is usually nonexponential,
and $\Gamma_{\rm R}^{\rm{exp}}$ is the decay rate of the exponentially decaying term.
As we are interested in the flux fluctuations at the Rabi frequency,
contributions from other sources are to be separated out.


The quasistatic noise, which results in $A_{\rm st}(t)$ in Eq.~(\ref{RabiDecay}),
is attributed to the fluctuations of the time-averaged values of $\delta \varepsilon (t)$
and $\delta \Delta (t)$ during a single decoherence measurement trial.
The variances of the quasistatic flux noise, $\sigma_{\delta \varepsilon}^2$, and the $\Delta$ noise, $\sigma_{\delta\Delta}^2$,
are determined from the result of free-induction-decay (FID) measurements,~\cite{supRabi}
where we find $\sigma^2_{\delta \varepsilon} \gg \sigma^2_{\delta \Delta}$.
To evaluate the decay envelope $A_{\rm st}(t)$ due to the quasistatic flux noise,
we numerically calculate the time evolution of the density matrix of the qubit $\rho_{\rm qubit}(t)$
under $\mathcal{H}_{\rm pc}.$


The exponentially decaying component of the envelope is caused by the
fluctuations at $\omega_{01}$ and $\Omega_{\rm R}$,
and the rate is written as~\cite{Ithier05PRB}
\begin{eqnarray}\label{G2Rabi}
    \Gamma_{\rm R}^{\rm{exp}} & = & \frac{(3-\cos ^2 \zeta )\Gamma_1}{4} + \Gamma_{\Omega_{\rm R}},
\end{eqnarray}
where
\begin{eqnarray}\label{G1}
    \Gamma_1 = \frac{2\pi}{\hbar^2}\sum_{\lambda}S_{\lambda}(\omega_{01})
    \left |  \left \langle 1 \left | \frac{\partial \mathcal{H}_{\rm pc}}{\partial \lambda}\right |0 \right \rangle \right |^2
\end{eqnarray}
and
\begin{multline}\label{GRfR}
    \Gamma_{\Omega_{\rm R}} =  
     \sin ^2 \zeta\, \frac{\pi}{2\hbar^2}[(2I_{\rm p}\Phi_0)^2 S_{n_{\phi}}(\Omega_{\rm R})\cos^2\eta \\
     + \hbar^2 S_{\Delta}(\Omega_{\rm R})\sin^2\eta].
\end{multline}
Here $\zeta = \arccos (\delta \omega_{\rm mw}/\Omega_{\rm R})$, $\eta = \arctan (\Delta/\varepsilon)$,
$\delta \omega_{\rm mw} \equiv \omega_{\rm mw} -\omega_{01}$ is the detuning frequency,
and $S_{\lambda}(\omega) = \frac{1}{2\pi}\int_{-\infty}^{\infty}d\tau\langle \delta \lambda (t)\lambda (t+\tau)\rangle \exp (-i\omega \tau)$
denotes the PSD of a fluctuating parameter $\delta \lambda$
such as flux, charge, and critical current of the Josephson junctions.
$\Gamma_1$ is the rate of the energy relaxation induced by the fluctuations at $\omega_{01}$
and can be independently measured as the decay rate of the qubit population after a $\pi$-pulse excitation.~\cite{supRabi}
Strictly speaking, the first term in Eq.~(\ref{G2Rabi}) is written as $[\Gamma_1'+2\Gamma_1+(\Gamma_1'-2\Gamma_1)\cos^2\zeta]/4$,~\cite{Ithier05PRB}
where $\Gamma_1'$ is the average of the energy relaxation rates at $\omega_{01} \pm \Omega_{\rm R}$
and is usually close to $\Gamma_1$.
$\Gamma_{\Omega_{\rm R}}$ is the decay rate due to fluctuations at $\Omega_{\rm R}$.
Therefore, by analyzing experimental results using Eqs.~(\ref{RabiDecay})--(\ref{GRfR}),
$S_{n_{\phi}}(\Omega_{\rm R})$ and $S_{\Delta}(\Omega_{\rm R})$ can be evaluated from the Rabi oscillation measurements
at $\varepsilon \ll \Delta$ and $\varepsilon \approx \Delta$.

We need to pay attention to the drive-induced frequency shift of the qubit
in the Rabi oscillation measurements under strong driving.
We resort to numerical calculations to study the shift of the resonant frequency $\delta \omega$
as a function of $\varepsilon_{\rm{mw}}$.
At each $\varepsilon_{\rm{mw}}$,
$\Omega_{\rm R}$ is calculated as a function of $\omega_{\rm mw}$ and fitted with an analytic form,
\begin{eqnarray}\label{fRfull}
  \Omega_{\rm R} =
  \sqrt{\left (\frac{\Delta}{2}\frac{\varepsilon_{\rm{mw}}}{\omega_{01}}\right )^{\!2}  
  + \left [ \omega_{\rm{mw}}-\left (\omega_{01} + \delta \omega \right ) \right ] ^{2}}.
\end{eqnarray}
The first term, $(\frac{\Delta}{2}\frac{\varepsilon_{\rm{mw}}}{\omega_{01}})^2 \equiv \Omega_{\rm R0}^2$,
is the square of the Rabi frequency at the new resonance condition ($\omega_{\rm{mw}} = \omega_{01} + \delta \omega$),
and the second term is the square of the detuning from the resonance.
For simplicity, we use the linear approximation, $\Omega_{\rm R0} \propto \varepsilon_{\rm mw}/\omega_{01}$.
This approximation is numerically validated within the range of parameters
$\Omega_{\rm R0}$ and $\varepsilon$ in most cases in the present study.

\begin{figure}
\includegraphics[width=\linewidth]{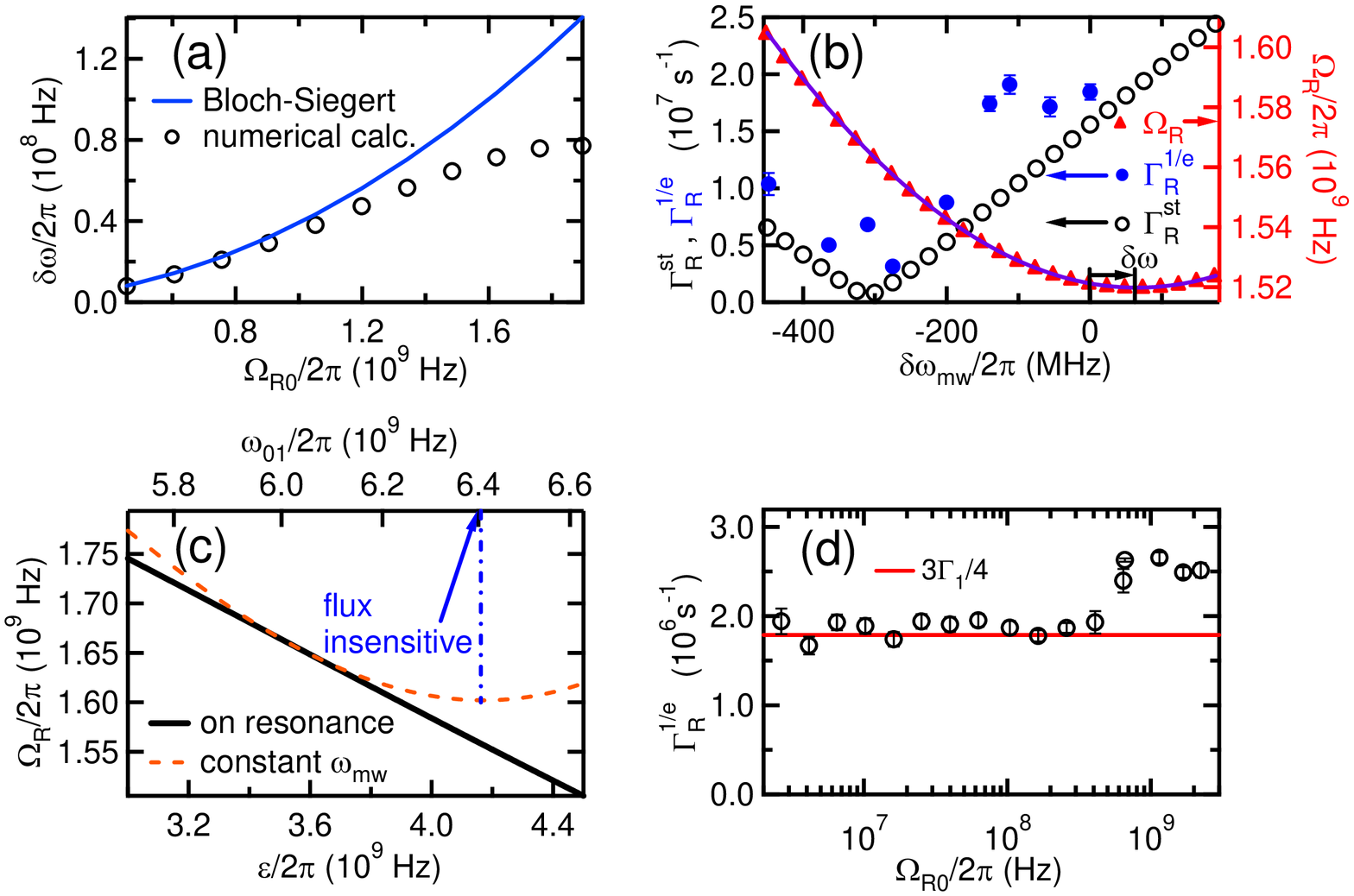}
\caption{(Color online) (a)~Numerically calculated shift of the resonant frequency $\delta \omega$ (black open circles)
and the Bloch--Siegert shift $\delta \omega_{\rm BS}$ (blue line).
(b)~Numerically calculated decay rate $\Gamma_{\rm R}^{\rm st}$ (black open circles) and
Rabi frequency $\Omega_{\rm R}$ (red solid triangles)
as functions of the detuning $\delta \omega_{\rm{mw}}$ from $\omega_{01}$.
The purple solid line is a fit based on Eq.~(\ref{fRfull}).
The measured 1/$e$ decay rates $\Gamma_{\rm R}^{1/e}$ at $\varepsilon/2\pi =$ 4.16~GHz
for the range of Rabi frequencies $\Omega_{\rm R}/2\pi$ between 1.5 and 1.6~GHz (blue solid circles) are also plotted.
(c)~Calculated Rabi frequency $\Omega_{\rm R}$, based on Eq.~(\ref{fRfull}), as a function of $\varepsilon$
for the cases (i)~$\omega_{\rm mw} = \omega_{01} + \delta \omega$ (black solid line) and
(ii)~$\omega_{\rm mw}/2\pi = 6.1$~GHz (red dashed line).
The upper axis indicates $\omega_{01}$, corresponding to $\varepsilon$ in the bottom axis.
(d)~The measured $1/e$ decay rate of the Rabi oscillations, $\Gamma_{\rm R}^{1/e}$,
at $\varepsilon = 0$ and as a function of $\Omega_{\rm R0}$.
The red solid line indicates $\frac{3}{4}\Gamma_1$ obtained independently.
}
\label{GRfR1p5}
\end{figure}

In Fig.~\ref{GRfR1p5}(a), $\delta \omega$ as a function of $\Omega_{\rm R0}$ is plotted
together with the well-known Bloch--Siegert shift,~\cite{Bloch45PR, TannoudjiAPinter}
$\delta \omega_{\rm BS} = \frac{1}{4}\frac{\Omega_{\rm R0}^2}{\omega_{01}}$, obtained from the second-order perturbation theory.
Fixed parameters for the calculation are
$\Delta /2\pi = 4.869$ and $\varepsilon /2\pi = 4.154$~GHz ($\omega_{01}/2\pi = 6.400$~GHz).
We find that $\delta \omega_{\rm BS}$ overestimates $\delta \omega$
when $\Omega_{\rm R0}/2\pi \gtrsim 800$~MHz.
The deviation from the Bloch--Siegert shift is due to the component of the ac flux drive
that is parallel to the qubit's energy eigenbasis;
this component is not averaged out when $\Omega_{\rm R}$ is comparable to $\omega_{\rm mw}$.

We next calculate the decay of Rabi oscillations
due to quasistatic flux noise~\cite{supRabi}
and examine its dependence on $\delta \omega_{\rm{mw}}$.
In Fig.~$\ref{GRfR1p5}$(b), $\Omega_{\rm R}$ and the decay rate
$\Gamma_{\rm R}^{\rm st}$, defined as the inverse of the $1/e$ decay time,
are plotted as functions of $\delta \omega_{\rm mw}$.
Fixed parameters, $\varepsilon_{\rm mw}/2\pi = 4.100$~GHz and
$\sigma_{\delta \varepsilon}/2\pi = 27.8$~MHz, are chosen.
Interestingly, neither the minimum of $\Omega_{\rm R}$ nor that of $\Gamma_{\rm R}^{\rm st}$ is located at $\delta \omega_{\rm mw} = 0$,
but at $\delta \omega_{\rm mw}/2\pi =$ 66.5~MHz for $\Omega_{\rm R}$ and at $\delta \omega_{\rm mw}/2\pi = -311$~MHz for $\Gamma_{\rm R}^{\rm st}$.
For the ac flux drive, the frequency offset that minimizes the Rabi frequency
is a consequence of the amplitude-dependent frequency shift $\delta \omega$, as can be observed in Eq.~(\ref{fRfull}).
Since the fluctuations of $\Omega_{\rm R}$ causes the decay of Rabi oscillations,~\cite{Solomon59PRL}
the minimum of $\Gamma_{\rm R}^{\rm st}$ is understood by
considering the flux sensitivity of $\Omega_{\rm R}$ which is expressed as
\begin{eqnarray}
  \frac{\partial \Omega_{\rm R}}{\partial \varepsilon} = \frac{-\varepsilon}{\omega_{01}\Omega_{\rm R}}
  \left [\omega_{\rm{mw}}-\left (\omega_{01}+\delta \omega - \frac{\Omega_{\rm R0}^2}{\omega_{01}}\right )\right ].
\end{eqnarray}
The condition, $\partial \Omega_{\rm R}/\partial \varepsilon = 0$, is satisfied when $\varepsilon = 0$ or
$\delta \omega_{\rm{mw}} = \delta \omega - \Omega_{\rm{R0}}^2/\omega_{01}$.
For $\Omega_{\rm{R0}}/2\pi$ = 1.52~GHz and $\omega_{01}/2\pi$ = 6.400~GHz,
the latter condition is calculated to be $\delta \omega_{\rm mw}/2\pi = -295$~MHz,
slightly different from the minimum of $\Gamma_{\rm R}^{\rm st}$ seen in Fig.~$\ref{GRfR1p5}$(b).
The difference is due to the deviation from the linear approximation in Eq.~(\ref{fRfull}),
$\Omega_{\rm R0} \propto \varepsilon_{\rm mw}/\omega_{01}$.
Figure~\ref{GRfR1p5}(c) shows the calculation of $\Omega_{\rm R}$ as a function of $\varepsilon$,
based on Eq.~(\ref{fRfull}).
The Rabi frequency $\Omega_{\rm R0}$ at the shifted resonance decreases as $\varepsilon$ increases,
while $\Omega_{\rm R}$, for a fixed microwave frequency of $\omega_{\rm mw}/2\pi = 6.1$~GHz,
has a minimum of approximately $\omega_{01}/2\pi =$ 6.4~GHz.
Here in the first order,
$\Omega_{\rm R}$ is insensitive to the fluctuation of $\varepsilon$.



The experiments were performed with a sample fabricated by electron-beam lithography
 and shadow evaporation of Al films,
with a thickness of 13~nm for the first layer and 30~nm for the second,
 on an undoped Si substrate covered with a 300-nm-thick SiO$_2$ layer.~\cite{supRabi}
The qubit is a superconducting loop intersected by four Josephson junctions,
among which one is smaller than the others by a factor of 0.5, nominally.
The loop area is larger than that of flux qubits that we previously used,~\cite{Fum06PRL}
yielding a large mutual inductance between the qubit and the microwave line ($1.2 \, \mathrm{pH}$)
and facilitating strong driving.

We first measured $\omega_{01}$ as a function of $\varepsilon$ and determined the qubit parameters.
A 1-$\mathrm{\mu s}$ microwave pulse is applied to the qubit, followed by a bias current pulse of the readout SQUID (readout pulse).
When the microwave frequency hits a transition of the qubit,
 the excitation is detected as a change in the SQUID switching probability $P_{\rm{sw}}$.
The flux qubit under study was cooled twice in between, up to room temperature with a thermal cycling.
We noticed that $\Delta$ decreased by 1$\%$ after the thermal cycling:
$\Delta/2\pi = 4.87$~GHz during the first cooldown
and $\Delta/2\pi = 4.82$~GHz during the second.
$I_{\rm{p}} = 235$~nA was the same for both cooldowns.
Unless explicitly mentioned below, we present the data from the first cooldown.

\begin{figure}
\includegraphics[width=\linewidth]{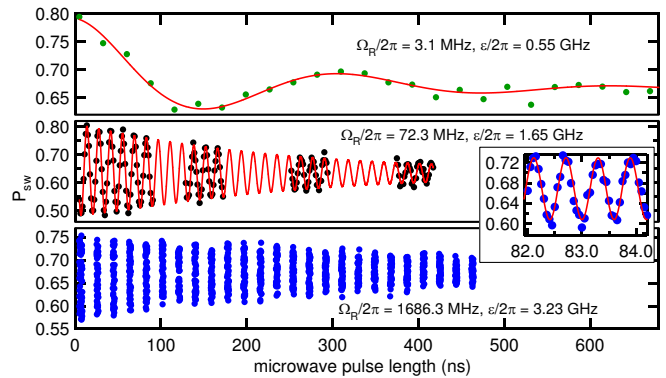}
\caption{(Color online) Rabi oscillation curves with different Rabi frequencies $\Omega_{\rm R}$
measured at different static flux bias $\varepsilon$.
At each $\Omega_{\rm R}$,
$\delta \omega_{\rm{mw}}$ is chosen to minimize dephasing due to quasistatic flux noise.
The red lines are the fitting curves.~\cite{supRabi}
In the measurements shown in the middle and bottom panels,
only parts of the oscillations are monitored
so that we can save measurement time while the envelopes of Rabi oscillations are captured.
The inset is a magnification of the data in the bottom panel together with the fitting curve.  
}
\label{Rabis}
\end{figure}

In the Rabi oscillation measurements, a microwave pulse is applied to the qubit followed by a readout pulse,
and $P_{\rm sw}$ as a function of the microwave pulse length is measured.
First, we measure the Rabi oscillation decay at $\varepsilon = 0$,
where the quasistatic noise contribution is negligible.
Figure~\ref{GRfR1p5}(d) shows the measured $1/e$ decay rate of the Rabi oscillations
$\Gamma_{\rm R}^{1/e}$ as a function of $\Omega_{\rm R0}$.
For $\Omega_{\rm R0}/2\pi$ up to 400~MHz,
$\Gamma_{\rm R}^{1/e}$ is approximately $3\Gamma_1/4$, limited by the energy relaxation,
and $S_{\Delta}(\Omega_{\rm R0})$ is negligible.
For $\Omega_{\rm R0}/2\pi$ from 600~MHz to 2.2~GHz,
$\Gamma_{\rm R}^{1/e} > 3\Gamma_1/4$.
A possible origin of this additional decoherence is fluctuations of $\varepsilon_{\rm mw}$,
$\delta\varepsilon_{\rm mw}$:
$\Omega_{\rm R0}$ is first order sensitive to $\delta\varepsilon_{\rm mw}$,
which is reported to be proportional to $\varepsilon_{\rm mw}$ itself.~\cite{Gustavsson12PRLRE}
Next, the decay for the case $\varepsilon \approx \Delta$ is studied.
To observe the contribution from quasistatic flux noise,
the Rabi oscillation decay as a function of $\omega_{\rm mw}$ is measured,
where the contribution from the other sources is expected to be almost constant.
Figure~\ref{GRfR1p5}(b) shows $\Gamma_{\rm R}^{1/e}$
at $\varepsilon/2\pi =$ 4.16~GHz as a function of $\delta \omega_{\rm{mw}}$
while keeping $\Omega_{\rm R}/2\pi$ between 1.5 and 1.6~GHz.
Besides the offset and scatter, the trend of $\Gamma_{\rm R}^{1/e}$ agrees with
that of the simulated $\Gamma_{\rm R}^{\rm st}$.
This result indicates that numerical calculation properly evaluates $\delta \omega_{\rm mw}$ minimizing $\Gamma_{\rm R}^{\rm st}$.
Finally, the decay for the case $\varepsilon \approx \Delta$
as a function of $\varepsilon_{\rm{mw}}$,
covering a wide range of $\Omega_{\rm R}$, is measured (Fig.~\ref{Rabis}).
At each $\Omega_{\rm R}$,
$\delta \omega_{\rm{mw}}$ is chosen to minimize dephasing due to quasistatic flux noise,
which is numerically calculated as $A_{\rm st}(t)$ in Eq.~(\ref{RabiDecay}).
After dividing $A_{\rm env}(t)$ by $A_{\rm st}(t)$ in Eq.~(\ref{RabiDecay})
and subtracting the decay rates obtained by $\Gamma_1$ and $S_{\Delta}(\Omega_{\rm R})$
from $\Gamma_{\rm R}^{\rm exp}$ using Eqs.~(\ref{G2Rabi})--(\ref{GRfR}),
$S_{n_{\phi}}(\Omega_{\rm R})$ is extracted.~\cite{supRabi}
Parameters in calculations and measurements are summarized in Table I.

\begin{table}
\caption{Parameters in calculations and measurements in units of GHz.
In the first column, cal: $\delta\omega(\Omega_{\rm R0})$
stands for the calculation to study the shift of the resonant frequency,
and cal: $\Gamma_{\rm R}^{\rm st}(\delta\omega_{\rm mw})$
stands for the calculation to study the decay of Rabi oscillations due to quasistatic flux noise.
``Optimal" in the last column means that at each $\varepsilon_{\rm mw}$,
$\omega_{\rm mw}$ is chosen to minimize dephasing due to quasistatic flux noise.
}
\begin{tabular}{c c c c c}
    \hline
    \hline
     & $\Delta/2\pi$ & $\varepsilon/2\pi$ & $\varepsilon_{\rm mw}/2\pi$ & $\delta\omega_{\rm mw}/2\pi$\\
    \hline
    cal: $\delta\omega(\Omega_{\rm R0})$ & 4.869 & 4.154 & 1.2 -- 5.0 & $-0.02$ -- 0.12\\
    cal: $\Gamma_{\rm R}^{\rm st}(\delta\omega_{\rm mw})$ & 4.869 & 4.154 & 4.100 & $-0.45$ -- 0.175\\
    Cooldown1 & 4.87 & 0, 4.16 & 0.005 -- 4.5 & optimal\\
    Cooldown2 & 4.82 & 0.55 -- 3.23 & 0.02 -- 0.16 & optimal\\
    \hline
    \hline
\end{tabular}
\label{cal_meas}
\end{table}

\begin{figure}
\includegraphics[width=\linewidth]{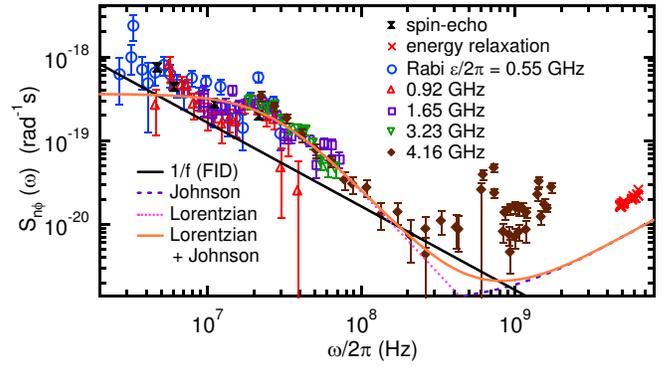}
\caption{(Color online) Power spectrum density of flux fluctuations $S_{n_{\phi}}(\omega)$
 extracted from the Rabi oscillation measurements in the first ($\varepsilon/2\pi =$ 4.16~GHz)
 and second cooldowns.
 The PSDs obtained from the spin-echo and energy relaxation measurements in the second cooldown are also plotted.
 The black solid line is the 1/$f$ spectrum extrapolated from the FID measurements in the second cooldown.
 The purple dashed line is the estimated Johnson noise from a 50~$\Omega$ microwave line
 coupled to the qubit by a mutual inductance of 1.2~pH and nominally cooled to 35~mK.
 The pink dotted line is a Lorentzian,
 $S_{n_{\phi}}^{\rm model}(\omega) = S_{\rm{h}} \omega_{\rm{w}}^2/(\omega^2+\omega_{\rm{w}}^2)$,
 and the orange solid line is the sum of the Lorentzian and the Johnson noise.
 Here the parameters are $S_{\rm{h}} = 3.6 \times 10^{-19}\, \mathrm{\rm rad^{-1}\,s}$ and
 $\omega_{\rm{w}}/2\pi = 2.7\times 10^7\, \mathrm{Hz}$.
}
\label{SnfWlor}
\end{figure}

The PSD of flux fluctuations $S_{n_{\phi}}(\omega)$, evaluated from the Rabi oscillation measurements
in the first and second cooldowns,
and PSDs from the spin-echo and energy relaxation measurements~\cite{supRabi}
in the second cooldown are plotted in Fig.$~\ref{SnfWlor}$.
The 1/$f$ spectrum extrapolated from the FID measurements in the second cooldown,
$S_{n_{\phi}}(\omega) = (3.2\times 10^{-6})^2/\omega$,~\cite{supRabi} is also plotted.
Several points are worth mentioning:
(i)~$S_{n_{\phi}}(\omega)$ from the Rabi oscillation measurements in the first and second cooldowns is consistent.
(ii)~$S_{n_{\phi}}(\omega)$ from the spin-echo measurements is consistent with that from the Rabi-oscillation measurements.
(iii)~$S_{n_{\phi}}(\omega)$ from the energy relaxation measurements is 2.5 times larger than
expected for the decay into a 50~$\Omega$ microwave line
coupled to the qubit by a mutual inductance of 1.2 pH and nominally cooled to 35 mK.
(iv)~There can be an additional decoherence induced by strong driving as observed in Fig.$~\ref{GRfR1p5}$(d),
so, it is not surprising to see the increased and scattered $S_{n_{\phi}}(\omega)$
from the Rabi oscillation measurements above 300~MHz.
These data points should be considered as the upper limit of the noise.
(v)~$S_{n_{\phi}}(\omega)$ from the Rabi oscillation measurements is roughly parallel to the 1/$f$ spectrum
extrapolated from the FID measurements but is larger in general and has more structures:
the deviation is largest at 25~MHz, and the slope at approximately 100~MHz is steeper than 1/$f$.
(vi)~$S_{n_{\phi}}(\omega)$ around 300~MHz is approximately $10^{-20}\,\rm rad^{-1}s$,
which is (number) orders of magnitude smaller than those reported,~\cite{Slichter12PRL}
demonstrating that the noise level is not very far from the extrapolation of the $1/f$ spectrum,
even at such high frequencies.

We consider localized electron spins on the surface of the superconducting loop~\cite{Sendelbach08PRL, Bluhm09PRL, Faoro08PRL, Choi09PRL}
as a possible cause of the PSD of flux fluctuations.
The total number of electron spins is estimated to be $9\times 10^6$,
adopting the reported surface spin density of $5\times10^{17}\,\mathrm{m^{-2}}$~\cite{Sendelbach08PRL}
and the total surface area of $\sim 19\,\mathrm{\mu m^2}$
considering both the top and bottom surfaces of the superconducting loop;
the loop of the qubit has a $4.8\times 6.8 \,\mathrm{\mu m^2}$ rectangular shape,
and the line width is 400~nm.

The magnetic field perpendicularly applied to the qubit loop was approximately 2~G, and
screening due to the superconducting film leads to a variation of the field;
the magnetic field at the top and bottom surfaces of the loop is shielded,
while the field at the edge of the loop is doubled.
Considering that the corresponding Zeeman splitting, at most $h\times 11$~MHz,
is much smaller than the thermal energy at 35 mK,
the electron spins are expected to be oriented randomly.
Because of the broad spectrum of the Zeeman splitting,
a clear signal from the electron spin resonance is not expected in $S_{n_{\phi}}(\omega)$.

We next consider the case where each electron spin generates a random telegraph signal (RTS).
The PSD of flux RTSs generated by $N$ independent electron spins is written
as a sum of Lorentzians:~\cite{Cywinski08PRB}
\begin{eqnarray}\label{RTSs}
    S_{n_{\phi}}^{\rm RTS}(\omega) & = &  \frac{N}{3}n_{\phi e}^2 \sum_{i = 1}^N \frac{1}{\pi}\frac{2\gamma_i}{\omega^2+4\gamma_i^2},
\end{eqnarray}
where $\gamma_i$ is the mean rate of transition per second between two states of the $i$th electron spin
and $n_{\phi e}$ is a normalized flux through the qubit loop in units of $\Phi_0$.
Here $n_{\phi e}$ is induced by an electron spin parallel to the magnetic field
generated by the persistent current in the qubit loop.
For simplicity, we use a constant normalized flux $n_{\phi e} = 1.3 \times 10^{-8}$.~\cite{supRabi} 

In the case of a 1/$f$ spectrum, the distribution function of $\gamma$ is expressed as $g(\gamma)\propto 1/\gamma$.
On the other hand, we speculate that the steep slope at approximately 100~MHz in $S_{n_{\phi}}(\omega)$ is a part of a Lorentzian,
$S_{n_{\phi}}^{\rm model}(\omega) = S_{\rm{h}} \omega_{\rm{w}}^2/(\omega^2+\omega_{\rm{w}}^2)$,
where $S_{\rm{h}}$ and $\omega_{\rm{w}}$ are the height and the width of the Lorentzian peak, respectively.
In Fig.~$\ref{SnfWlor}$, an example of $S_{n_{\phi}}^{\rm model}(\omega)$ is also plotted.
Here we chose $S_{\rm{h}} = 3.6 \times 10^{-19}\, \mathrm{rad^{-1}\,s}$
and $\omega_{\rm{w}}/2\pi = 2.7\times 10^7\, \mathrm{Hz}$,
and $S_{n_{\phi}}^{\rm model}(\omega)$ amounts to the PSD generated by
$3.6 \times 10^6$ independent electron spins with the same transition rate of $\gamma = 8.5\times 10^7\,\rm{s}^{-1}$.
The number of electron spins corresponds to approximately 40$\%$ of the total surface spins.
The number would be smaller in the case where electron spins form ferromagnetic clusters
and the spins in each cluster flip simultaneously.~\cite{Sendelbach09PRL}
The rest of the surface spins may form a 1/$f$ spectrum up to a few megahertz,
where $S_{n_{\phi}}^{\rm model}(\omega)$ deviates from $S_{n_{\phi}}(\Omega_{\rm R})$.
To further investigate the origin of the flux noise,
a systematic study of the PSD in the high frequency domain is required.

In conclusion, we have evaluated the PSD of flux fluctuations in a superconducting flux qubit
by measuring the decay of Rabi oscillations.
The measured Rabi frequency ranges from 2.7~MHz to 1.7~GHz,
close to the qubit's level splitting of 4.8~GHz.
The observed PSD decreases up to 300~MHz,
where the PSD is approximately $10^{-20}\,\rm rad^{-1}s$,
not very far from the $1/f$ spectrum extrapolated from the FID measurements.

We are grateful to L. Ioffe, L. Faoro, and P.-M. Billangeon for their valuable discussions.
This study was supported by the Grant-in-Aid for Scientific Research Program for
Quantum Cybernetics of the Ministry of Education, Culture, Sports, Science, and Technology (MEXT), Japan,
Funding Program for World-Leading Innovative R$\&$D on Science and Technology (FIRST),
and the NICT Commissioned Research.


\end{document}